\documentclass[aps,twocolumn,floatfix,showpacs,superscriptaddress]{revtex4-1}
\pdfoutput=1
\usepackage{graphicx}
\usepackage{amsmath}
\usepackage[colorlinks=true, citecolor=blue]{hyperref}
\usepackage{amssymb}
\usepackage{dsfont}
\usepackage{bm}
\usepackage{bbm}
\usepackage{color}

\DeclareMathOperator{\de}{d\!}

\newcommand{\bra}[1]{\langle #1|}
\newcommand{\ket}[1]{|#1\rangle}
\newcommand{\bracket}[2]{\langle #1|#2\rangle}

\newcommand{\pd}{\partial}

\begin{document}

\title{Braiding of non-Abelian anyons using pairwise interactions}
\author{M. Burrello}
\author{B. van Heck}
\affiliation{Instituut-Lorentz, Universiteit Leiden, P.O. Box 9506, 2300 RA Leiden, The Netherlands}
\author{A. R. Akhmerov}
\affiliation{Instituut-Lorentz, Universiteit Leiden, P.O. Box 9506, 2300 RA Leiden, The Netherlands}
\affiliation{Department of Physics, Harvard University, Cambridge, MA 02138}

\begin{abstract}
  The common approach to topological quantum computation is to implement
  quantum gates by adiabatically moving non-Abelian anyons around each
  other. Here we present an alternative perspective based on the possibility of
  realizing the exchange (braiding) operators of anyons by adiabatically
  varying pairwise interactions between them rather than their positions. We
  analyze a system composed by four anyons whose couplings define a T-junction
  and we show that the braiding operator of two of them can be obtained through
  a particular adiabatic cycle in the space of the coupling parameters. We also
  discuss how to couple this scheme with anyonic chains in order to recover the
  topological protection.
\end{abstract}

\pacs{03.67.Lx, 03.65.Vf, 05.30.Pr}

\maketitle

\section{Introduction}

\begin{figure}[tb] 
\includegraphics[width=0.9\linewidth]{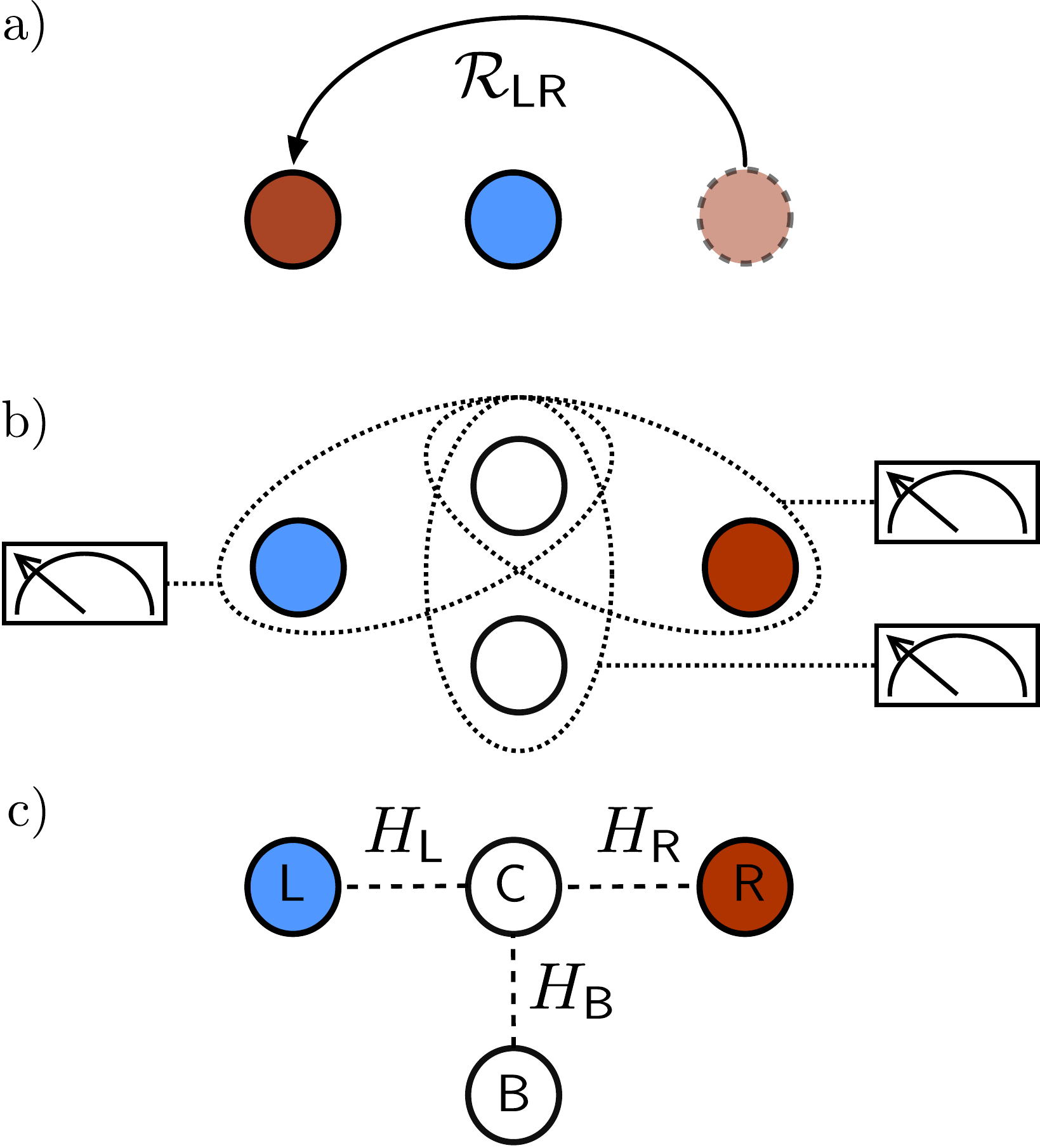}
\caption{Different ways to braid quasiparticles in topological quantum
  computation. Panel (a): the original scheme for braiding, where quantum gates
  are obtained by moving the non-Abelian quasiparticles (red and blue dots) one
  around the other. Panel (b): measurement-only TQC, in which ancillary anyons
  are added to the system (white dot), and quantum gates are obtained as a
  sequence of non-demolition pairwise measurements (represented by the dashed
  ellipses) which induce teleportation of the computational anyons through the
  ancillary ones. Panel (c): the interaction-based braiding, which makes use of
  the interaction between computational and ancillary anyons in a T-junction
  geometry.}
\label{fig_layout}
\end{figure}

The purpose of topological quantum computation (TQC) is to realize a reliable
quantum computer, exploiting the existence of special quasiparticles, known as
non-Abelian anyons, in certain exotic condensed matter systems
\cite{kitaev03,nayak08}. The presence of several such particles gives rise to
degenerate ground states which cannot be distinguished by local
measurements. The ground state manifold is then adopted as the computational
space, and quantum gates can be performed by braiding (exchanging the positions
of the anyons), as shown in Fig.~\ref{fig_layout}a). The resulting unitary
transformation of the wave function depends only on the order of the exchanges
and not on the details of their paths, thus these quantum gates are said to be
topologically protected. In the standard scheme of TQC \cite{nayak08}, there
are two main ingredients needed to implement braiding. First, it must be
possible to change the positions of anyons in such a way that the wave function
of the system always belongs to the space of the degenerate ground
states. Second, at all stages of the braiding the interactions between the
anyons used for the computation must be negligible in order to preserve the
degeneracy of the ground states and to avoid the presence of non-adiabatic
time-dependent phases. This requires the anyons to be well separated in space.

The possibility to realize braiding operators without moving the anyons was
then introduced by Bonderson, Freedman and Nayak in
Refs.~\cite{bonderson08,bonderson09}. In their scheme, the measurement-only
TQC, the braid operators are obtained as a result of a probabilistically
determined sequence of non-demolition measurements of the computational anyons
as shown in Fig.~\ref{fig_layout}b). This measurement would rely, for example,
on the non-Abelian edge state interferometry
\cite{fradkin98,dassarma05,stern06,bonderson06,bonderson08-2,bishara09}, which
has been actively developed both experimentally and theoretically
\cite{overbosch07,bishara08,rosenow08,willett09,willett10,kang11,simon11,clarke11,willett12}.

A different way to braid non-Abelian anyons without moving them around each
other has been theoretically developed in the case of Majorana fermions
appearing at the ends of one-dimensional topological superconductors
\cite{Kit01,Lut10,Ore10}. Initially, it was shown in Ref. \cite{Ali11} how
braids can be realized in wire networks by moving the Majorana fermions through
T-shaped junctions. In this case, the movement of the quasiparticles is
restricted to a quasi one-dimensional system, thus relaxing the limitation of
braiding to two dimensions. Subsequent proposals however have eliminated the
need to physically move the topological defects altogether, showing how the
same ground state transformations can be implemented using the mutual
interactions between Majoranas, controlling either tunnel couplings via gate
voltages \cite{Sau11} or capacitive couplings via magnetic fluxes
\cite{vanheck11}. Finally, in Ref. \cite{hal12} a general theory of adiabatic
manipulations of Majorana fermions in nanowires was formulated.  Unless we
allow for physically bending and rotating the wires, the minimal setup required
for the braid operation is a T-shaped configuration of nanowires where a
central Majorana is coupled to at least three neighbors. The evolution over a
path in parameter space results in the same non-Abelian Berry phase expected
after an exchange of two quasiparticles in real space.

In this paper, we aim to show that in a broad range of anyonic models braiding
is not only a property of the particle motion, but it is also encoded in the
many-body Hamiltonian of coupled anyons. We will show how it is possible to
engineer effective braidings by manipulating mutual couplings between
neighboring anyons, rather than their coordinates in space. The motion of
anyons is unnecessary also in measurement-only TQC, however our proposal is
different because the braid operation is performed in a deterministic manner
and does not rely on the procedure of anyon measurement.

The outline of the paper is the following. In Section \ref{Sec_Hamiltonian} we
present the minimal braiding setup, formed by four anyons in a T-shaped
junction, and we give an expression of the interaction Hamiltonian in terms of
the $\mathcal{F}$-matrices of a generic anyon model. In Sec. \ref{Sec_cycle} we
present in detail the adiabatic cycle in parameter space used to braid the
non-Abelian anyons, while in Sec. \ref{Sec_conclusions} we discuss how errors
affecting the adiabatic evolution can be reduced by embedding the braiding
junction in a bigger system of anyon chains and conclude.

\section{The T-junction}\label{Sec_Hamiltonian}

\begin{figure}[tb]
\includegraphics[width=\linewidth]{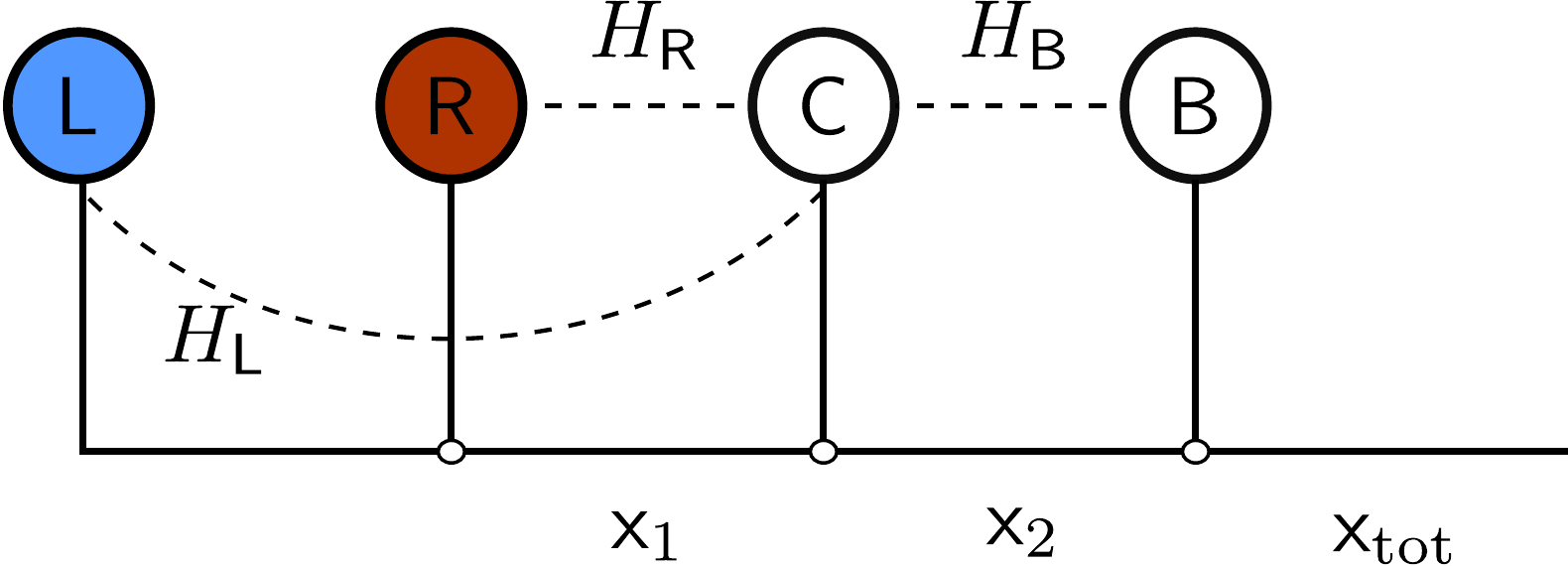}
\caption{Graphical representation of the T-junction system as a fusion tree of
  the four anyons, corresponding to the basis choice made in the text, see
  Eq. \eqref{basis}. Different sequences of the fusion outcomes $\mathsf{x}_1,
  \mathsf{x_2}, \mathsf{x}_\textrm{tot}$ define the basis states of the Hilbert
  space. Three Hamiltonians $H_\mathsf{L}, H_\mathsf{R}, H_\mathsf{B}$ describe
  the interaction between different pairs of anyons. In particular,
  $H_\mathsf{L}$ couples the anyons $\mathsf{L}$ and $\mathsf{C}$ which, in
  this basis, are not nearest neighbors.}
\label{fig_braider}
\end{figure}

We consider a system of four anyons with the same topological charge
$\textsf{t}$ in a T-junction geometry, with a central anyon (labeled
$\mathsf{t_C}$) coupled to other three (labeled $\mathsf{t_L}$, $\mathsf{t_R}$,
$\mathsf{t_B}$ for left, right and bottom), as shown in Fig. \ref{fig_layout}c
and Fig. \ref{fig_braider}. We assume that they have fusion rules
\begin{equation}\label{fusions}
\mathsf{t}\times\mathsf{t}=\sum_{i=1}^n \,\mathsf{f}_i
\end{equation}
with $\{\mathsf{f}_i\}$ the set of the $n$ possible fusion channels (see
Refs. \cite{kitaev06,nayak08,bondersonthesis,preskill} for introductions on
non-Abelian anyons and their fusion rules).

We also assume that the anyons do not move, and we focus on the pairwise
interactions between them. These interactions result in the fusion channels
$\mathsf{f}_i$ having different energies, so that the Hamiltonian can be written
as a sum of projectors onto different fusion outcomes. In the case of the
T-junction and given the fusion rule \eqref{fusions}, it takes the form
\begin{equation} \label{hamgen}
 H=-\sum_\mathsf{K} \sum_{i=1}^n \epsilon_{i,\mathsf{K}}\,\Pi^\mathsf{K}_i
\end{equation}
where $\mathsf{K}$ runs over $\{\mathsf{L}, \mathsf{B}, \mathsf{R}\}$ and
$\Pi^\mathsf{K}_{i}$ is the projector onto the states in which the anyon
$\mathsf{t_K}$ fuses with $\mathsf{t_C}$ into the $i$-th channel, with a
relative coupling $\epsilon_{i, \mathsf{K}}$. In order for braiding to work we require
that the interaction of each anyon with the central one favors an Abelian
channel $\mathsf{a_K}\in\{\mathsf{f}_i\}$, with a fusion energy $\epsilon_\mathsf{a,K} \equiv
\max\{\epsilon_{i,\mathsf{K}}\}$. This means that the anyons $\mathsf{C}$ and
$\mathsf{K}$ fusing in the $\mathsf{a_K}$ channel will be separated by an
excitation gap from all the other mutual fusion channels. In the following we
will assume that all the pairwise interactions favour the same fusion channel,
i.e. $\mathsf{a_L}=\mathsf{a_R}=\mathsf{a_B}=\mathsf{a}$, even though this
condition is not strictly necessary \cite{isingnote}.

To the purpose of implementing a braiding operator between anyons
$\mathsf{t_L}$ and $\mathsf{t_R}$ we require that all the pairwise interactions
$H_\mathsf{K}=\sum_{i=1}^n \epsilon_{i,\mathsf{K}}\,\Pi^\mathsf{K}_i$ in (\ref{hamgen}) can
be adiabatically switched off. In reality a single interaction $H_\mathsf{K}$
can not be totally switched off (even though it can be likely made
exponentially small), and we will relax this assumption in the
Sec.~\ref{Sec_chains}.

\subsection{Ground state degeneracy} \label{sec_degeneracy}

To prove that the Hamiltonian \eqref{hamgen} is of any use for TQC, we must
identify a degenerate manifold of its ground states, at least in some regions
of the parameter space spanned by the energies $\epsilon_{i,\mathsf{K}}$.

It has been shown that tunneling couplings between anyons lift completely the
topological degeneracy of the ground state \cite{bonderson09b}, and the
Hamiltonian \eqref{hamgen} makes no exception if all $\epsilon_\mathsf{a,K}$ are
non-zero. On the other hand, if all the couplings are zero, the ground state
manifold coincides with the whole Hilbert space of the anyon system. We focus
here on the intermediate domain between these two extreme cases, namely when
only a subspace of the full Hilbert space has its degeneracy left intact.

The Hamiltonian \eqref{hamgen} has an $n$-fold degenerate ground state when at
least one of the $H_\mathsf{K}$ is zero and one is non-zero. Let us consider
$H_\mathsf{L}=H_\mathsf{R}=0$, $\epsilon_\mathsf{a,B} > 0$. The two anyons
$\mathsf{L}$ and $\mathsf{R}$ are completely decoupled and share an arbitrary
topological charge $\mathsf{x}_1$ which may assume one of the $n$ different
values $\{\mathsf{f}_i\}$, while the anyons $\mathsf{B}$ and $\mathsf{C}$ fuse
into the Abelian channel $\mathsf{a}$. The total topological charge equals
$\mathsf{x}_\textrm{tot}=(\mathsf{t_R}\times
\mathsf{t_L})\times(\mathsf{t_B}\times \mathsf{t_C})=\mathsf{x}_1\times
\mathsf{a}$. Since $\mathsf{a}$ is Abelian, the fusion $\mathsf{x}_1\times
\mathsf{a}$ can only have one possible outcome, and additionally there cannot
be another charge $\mathsf{x}'_1$ such that $\mathsf{x}'_1\times \mathsf{a}$
has the same outcome. Therefore there exists a one-to-one mapping between the
charges $\mathsf{x}_1$ and $\mathsf{x}_\textrm{tot}$, implying that the ground
state wave function $\ket{\Psi}$ will generically be a superposition of $n$
orthogonal ground states $\Psi_i$ with total topological charge $\mathsf{f}_i
\times \mathsf{a}$, $\ket{\Psi}=\sum_i a_i \ket{\mathsf{f}_i\times \mathsf{a}}$.

When a second coupling, say $H_\mathsf{L}$, is also nonzero, the anyon
$\mathsf{L}$ fuses with $\mathsf{t_C}\times \mathsf{t_B}=\mathsf{a}$ and the three
have a total charge $\mathsf{t}\times \mathsf{a}$. The overall degeneracy cannot
change, since $\mathsf{t_R}\times(\mathsf{t_L}\times \mathsf{t_B} \times
\mathsf{t_C})=\mathsf{t}\times (\mathsf{t}\times \mathsf{a})=(\sum_i \mathsf{f}_i)\times
\mathsf{a}$, which again gives $n$ orthogonal states.

We conclude that if all the couplings $H_\mathsf{K}$ are neither \textit{on}
nor \textit{off} at the same time, the ground state of the Hamiltonian
\eqref{hamgen} has an $n$-fold degeneracy.

\begin{figure*}[t!]
\centering
\includegraphics[width=0.9\textwidth]{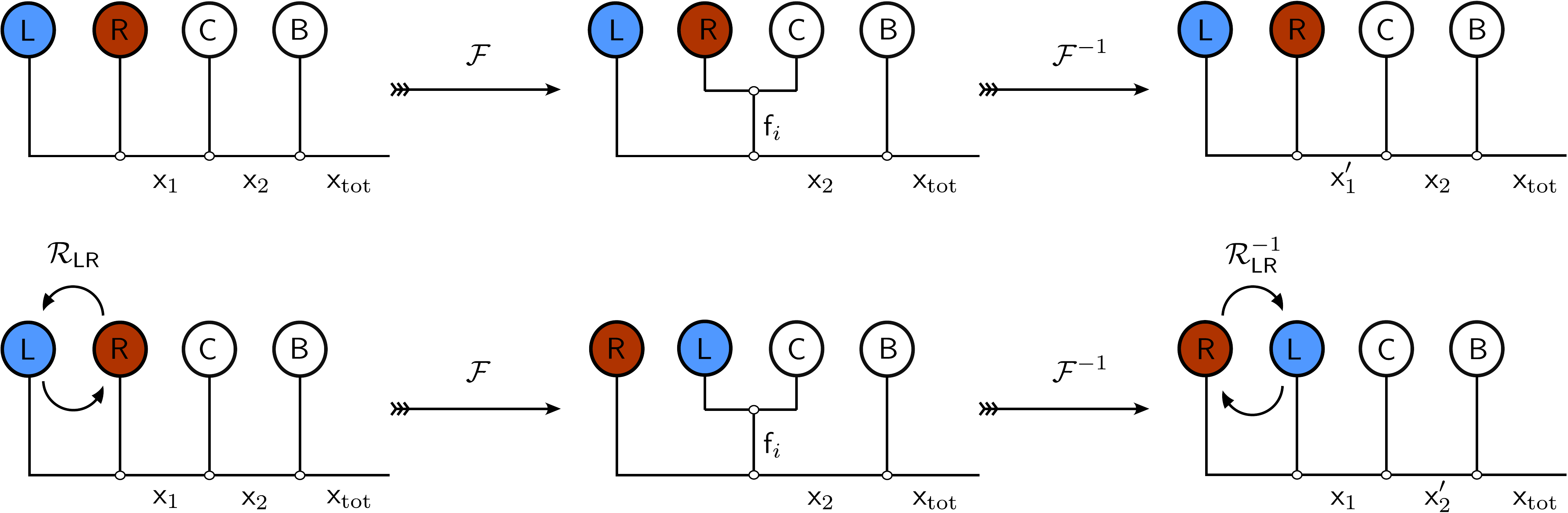}
\caption{Top: graphical representation of Eq. \eqref{PRC}. The fusion outcomes
  are explicitly written along the fusion tree. To write down the projectors
  $\Pi^\mathsf{R}_i$ in the basis of Fig. \ref{basis}, we need two
  $\mathcal{F}$-moves. A similar transformation, not shown, is needed to write
  down $\Pi^\mathsf{B}_i$, see Eq. \eqref{PCB}. Bottom: in the case of
  $\Pi^\mathsf{L}_i$, two braiding matrices $\mathcal{R}_\mathsf{LR}$ make their
  appearance in addition to the $\mathcal{F}$-moves. This introduces the
  braiding matrix $\mathcal{R}_\mathsf{LR}$ in the Hamiltonian of the
  $T$-junction.}
\label{fig_transformations}
\end{figure*}

\subsection{Projectors}\label{projectors}

In order to describe the wave function evolution in the $n$-fold degenerate
ground state subspace of \eqref{hamgen}, we need to write down the Hamiltonian
\eqref{hamgen} explicitly in a certain basis. To describe the evolution and the
eigenstates of this system we closely follow the methods used for the study of
anyon chains and lattices (see
e.g. \cite{feiguin07,trebst08,gils09,ardonne11,ludwig11}).

The different quantum states of a system of anyons can be specified by the
sequence of fusion outcomes along a certain fusion path. The choice of a fusion
path is equivalent to the choice of a basis in the Hilbert space. Once a fusion
path is chosen, the projector of two anyons on a given channel $\mathsf{f}_i$ is
represented by a simple diagonal matrix if the two anyons fuse directly
together along the path with outcome $\mathsf{f}_i$. Otherwise a projector must
be written via appropriate transformations called $\mathcal{F}$-matrices (see
e.g. \cite{nayak08,preskill,feiguin07}).  We choose the following fusion path
shown also in Fig. \ref{fig_braider}:
\begin{equation}\label{basis}
\left( \left( \left(\mathsf{t_L} \times \mathsf{t_R} \to \mathsf{x}_1 \right) \times \mathsf{t_C} \to \mathsf{x}_2 \right)\times \mathsf{t_B} \to \mathsf{x}_\textrm{tot}\right),
\end{equation}
with $\mathsf{x}_1,\mathsf{x}_2,\mathsf{x}_\textrm{tot}$ belonging to the sets of
possible fusion channels at each step of the fusion path. All states in the
Hilbert space can be written as $\ket{\mathsf{x}_1, \mathsf{x}_2,
  \mathsf{x}_\textrm{tot}}$. The basis (\ref{basis}) describes a path where
$\mathsf{t_L}$ and $\mathsf{t_R}$ are first fused with outcome $\mathsf{x}_1$,
then with $\mathsf{t_C}$ resulting in a second outcome $\mathsf{x}_2$, and
finally with the fourth anyon $\mathsf{t_B}$ to give
$\mathsf{x}_\textrm{tot}$. The latter is the total topological charge of the
system: subspaces of the Hilbert space corresponding to different
$\mathsf{x}_\textrm{tot}$ are decoupled. Adopting this basis we can now write
down explicitly all the terms appearing in the Hamiltonian \eqref{hamgen}. To
this purpose we consider different bases in which each operator has a diagonal
form, and then we move to the basis in Eq. \eqref{basis} using appropriate
basis transformations.

We start with $\Pi^\mathsf{B}_i$. The anyons $\mathsf{t_C}$ and $\mathsf{t_B}$ are
nearest neighbour, but they do not fuse directly together in our fusion path:
to write $\Pi_i^\mathsf{B}$ we must use the appropriate $\mathcal{F}$-matrices,
\begin{equation} \label{PCB}
\begin{split}
\left[ \Pi_{i}^\mathsf{B}(\mathsf{x_1},\mathsf{x}_\textrm{tot})\right]_{\mathsf{x}_2',\mathsf{x}_2} &=\sum_\mathsf{y} \left( \mathcal{F}^{\mathsf{x}_1\mathsf{t_Ct_B}}_{\mathsf{x}_\textrm{tot}} \right)^{-1}_{\mathsf{x}_2',\mathsf{f}_i}\,\delta_{\mathsf{f}_i,\mathsf{y}}\,\left( \mathcal{F}^{\mathsf{x}_1\mathsf{t_Ct_B}}_{\mathsf{x}_\textrm{tot}} \right)_{\mathsf{y},\mathsf{x}_2} =\\
&= \left( \mathcal{F}^{\mathsf{x}_1\mathsf{t_Ct_B}}_{\mathsf{x}_\textrm{tot}} \right)^{-1}_{\mathsf{x}_2',\mathsf{f}_i}\left( \mathcal{F}^{\mathsf{x}_1\mathsf{t_Ct_B}}_{\mathsf{x}_\textrm{tot}} \right)_{\mathsf{f}_i,\mathsf{x}_2}
\end{split}
\end{equation}
with $\mathsf{y}\,\in\,\{\mathsf{f_i}\}$ and $\mathsf{x}_2, \mathsf{x}'_2$ belonging to the set of fusion channels of three $\mathsf{t}$ anyons. As indicated on the left hand side of Eq. \eqref{PCB}, the matrix elements of the projector depend on indices $\mathsf{x_1}, \mathsf{x}_\textrm{tot}$. In a similar way we obtain for
$\Pi_i^\mathsf{R}$ the following form:
\begin{equation} \label{PRC}
 \left[ \Pi^\mathsf{R}_i(\mathsf{x_2})\right]_{\mathsf{x}_1',\mathsf{x}_1} = \left( \mathcal{F}^\mathsf{t_Lt_Rt_C}_{\mathsf{x}_2} \right)^{-1}_{\mathsf{x}_1',\mathsf{f}_i}\left( \mathcal{F}^\mathsf{t_Lt_Rt_C}_{\mathsf{x_2}} \right)_{\mathsf{f}_i,\mathsf{x}_1}
\end{equation}
with $\mathsf{x}_1, \mathsf{x}'_1\;\in\;\{\mathsf{f}_i\}$. The graphical representation of this equation is shown in the top panel of
Fig. \ref{fig_transformations}.

Unlike the two other cases, in the fusion tree of Fig. \ref{fig_braider} the
anyons $\mathsf{L}$ and $\mathsf{C}$ are not nearest neighbours in the chosen
basis. Since they would be nearest neighbors if $\mathsf{L}$ and $\mathsf{R}$
were interchanged, the transformation to a basis when they fuse directly
together includes a braiding matrix $\mathcal{R}_\mathsf{LR}$, as shown in the
bottom panel of Fig.~\ref{fig_transformations}. The particular braiding matrix
($\mathcal{R}_\mathsf{LR}$ or $\mathcal{R}^{-1}_\mathsf{LR}$) that appears in
this basis transformation depends on the real space positions of the anyons
 and on the microscopic details of the Hamiltonian. The two possible choices correspond to two mirror-symmetric anyon models \cite{bondersonthesis}. It is this
term that is responsible for the appearance of braiding during the adiabatic
Hamiltonian evolution. In particular mirroring the T-junction layout inverts the chirality of $\mathcal{R}$. As shown in the bottom
panel of Fig. \ref{fig_transformations}, the projector $\Pi^\mathsf{L}_i$ can
be obtained from $\Pi_i^\mathsf{R}$ via $\mathcal{R}_\mathsf{LR}$
\begin{equation}\label{symmetry1}
\mathcal{R}_\mathsf{LR}^{-1}\,\Pi^\mathsf{R}_i\mathcal{R}_\mathsf{LR}=\Pi^\mathsf{L}_i
\end{equation}
In the fusion basis \eqref{basis}, $\mathcal{R}_\mathsf{LR}$ is a diagonal
matrix and, explicitly, we have
\begin{equation}\label{PLC}
\begin{split}
\left[ \Pi^\mathsf{L}_i(\mathsf{x_2})\right]_{\mathsf{x}_1',\mathsf{x}_1} &= \left( \mathcal{R}_\mathsf{LR}^{-1}\right)_{\mathsf{x}'_1} \left( \Pi^\mathsf{R}_i\right)_{\mathsf{x}_1',\mathsf{x}_1} \left(\mathcal{R}_\mathsf{LR}\right)_{\mathsf{x}_1}=\\
&= \left(\mathcal{R}_\mathsf{LR}^{-1}\right)_{\mathsf{x}_1'}  \left( \mathcal{F}^\mathsf{t_Lt_Rt_C}_{\mathsf{x}_2} \right)^{-1}_{\mathsf{x}_1',\mathsf{f}_i}\left( \mathcal{F}^\mathsf{t_Lt_Rt_C}_{\mathsf{x}_2} \right)_{\mathsf{f}_i,\mathsf{x}_1} \left(\mathcal{R}_\mathsf{LR}\right)_{\mathsf{x}_1}.
\end{split}
\end{equation}
Knowing the $\mathcal{F}$-matrices of a given anyon model,
Eqs.~(\ref{PCB},\ref{PRC},\ref{PLC}) allow to write explicitly the four-anyon
Hamiltonian \eqref{hamgen}. In particular, we note that the braiding operator
$\mathcal{R}_\mathsf{LR}$ now appears explicitly in
\begin{equation}\label{hamL}
H_\mathsf{L}=\sum_i \epsilon_{i, \mathsf{L}}\,\Pi^\mathsf{L}_i=\sum_i \epsilon_{i, \mathsf{L}}\,\mathcal{R}^{-1}_\mathsf{LR}\,\Pi^\mathsf{R}_i\,\mathcal{R}_\mathsf{LR}.
\end{equation}

Before concluding this section, we point out that because the interactions are
local, the fusion product $\mathsf{t_B}\times\mathsf{t_C}$ cannot be affected by the
braiding of $R$ and $L$. The projectors $\Pi_i^\mathsf{B}$ and the braiding
operator $\mathcal{R}_\mathsf{LR}$ must therefore commute:
\begin{equation} \label{symmetry}
\Pi_i^\mathsf{B} \mathcal{R}_\mathsf{LR} =  \mathcal{R}_\mathsf{LR}\, \Pi_i^\mathsf{B}.
\end{equation}

\section{The adiabatic cycle}\label{Sec_cycle}

\begin{figure*}[t!]
 \centering
 \includegraphics[width=\textwidth]{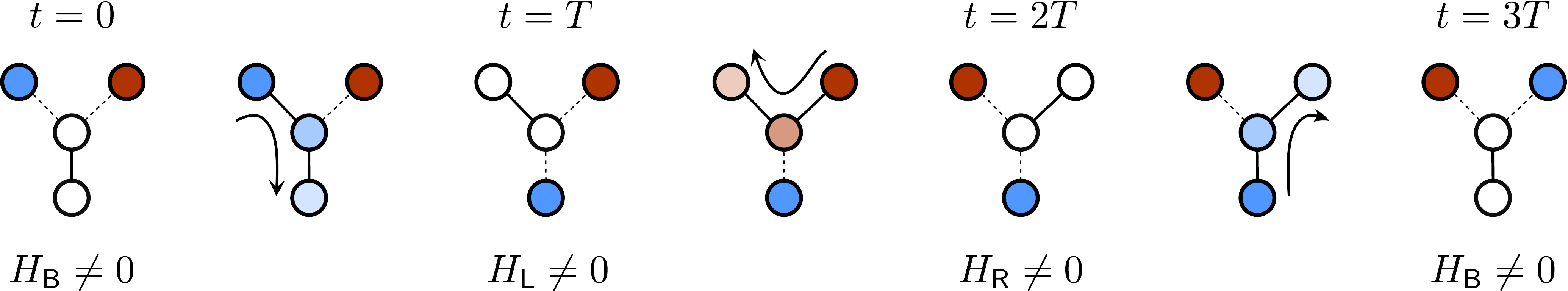}
 \caption{Illustration of the adiabatic cycle which reproduces the braiding
   operator $\mathcal{R}_\mathsf{LR}$ of two topological charges $\mathsf{t}$
   (red and blue circles) in a four anyon system. The cycle is divided in three
   steps of duration $T$. At the end of each step only one interaction
   $H_\mathsf{K}$ is on. The arrows follow the transfer of an unpaired
   topological charge $\mathsf{t}$ at intermediate stages, represented as the
   spreading of the colored circles over different anyons.} \label{fig_cycle}
\end{figure*}

In this section we show that the braiding of the anyons $\mathsf{R}$ and
$\mathsf{L}$ appears as a result of any closed path in parameter space starting
from a point where only $H_\mathsf{B} \neq 0$, and continuously passing through
the points where first only $H_\mathsf{L} \neq 0$, and finally only $H_\mathsf{R}
\neq 0$ in such a way that the degeneracy is always preserved. For the ease of
presentation we divide the path into three separate steps of duration $T$ such
that during each step one of $H_\mathsf{K}$ is turned on and one off. The time
evolution of the Hamiltonian along such a path is shown in
Fig.~\ref{fig_cycle}.

Let us consider the evolution of the ground state wave function $\ket{\Psi(t)}$ of
$H$ along this adiabatic cycle. The wave function can at any moment be written
as a superposition over states with different total topological charge
$\mathsf{x}_\textrm{tot}$,
\begin{equation}\label{gswf}
\ket{\Psi(t)}=\sum_{\mathsf{x}_\textrm{tot}}\,a_{\mathsf{x}_\textrm{tot}}\,\ket{\Psi_{\mathsf{x}_\textrm{tot}}(t)}.
\end{equation}
The states $\ket{\Psi_{\mathsf{x}_\textrm{tot}}(t)}$ define the $n$-fold ground
state manifold. The absolute values of the superposition coefficients
$a_{\mathsf{x}_\textrm{tot}}$ are conserved because the total topological
charge is a conserved quantity. This implies that the time evolution of the
ground state manifold is a diagonal operator in the basis given by
$\ket{\Psi_{\mathsf{x}_\textrm{tot}}(t)}$. Therefore, each term in the
superposition \eqref{gswf} can only acquire a phase, possibly dependent on
$\mathsf{x}_\textrm{tot}$, or in other words the Berry matrix is diagonal in
this basis. This allows us to follow the evolution of each
$\ket{\Psi_{\mathsf{x}_\textrm{tot}}(t)}$ independently from all other states.

We should note that the superposition \eqref{gswf} is only possible if other
anyons are present in the system other than $\mathsf{L}, \mathsf{R},
\mathsf{C}, \mathsf{B}$. We imagine that these anyons do not interact with the
T-junction while the adiabatic cycle is performed, so that their presence can be
ignored.

\begin{figure}[t!]
\includegraphics[width=\linewidth]{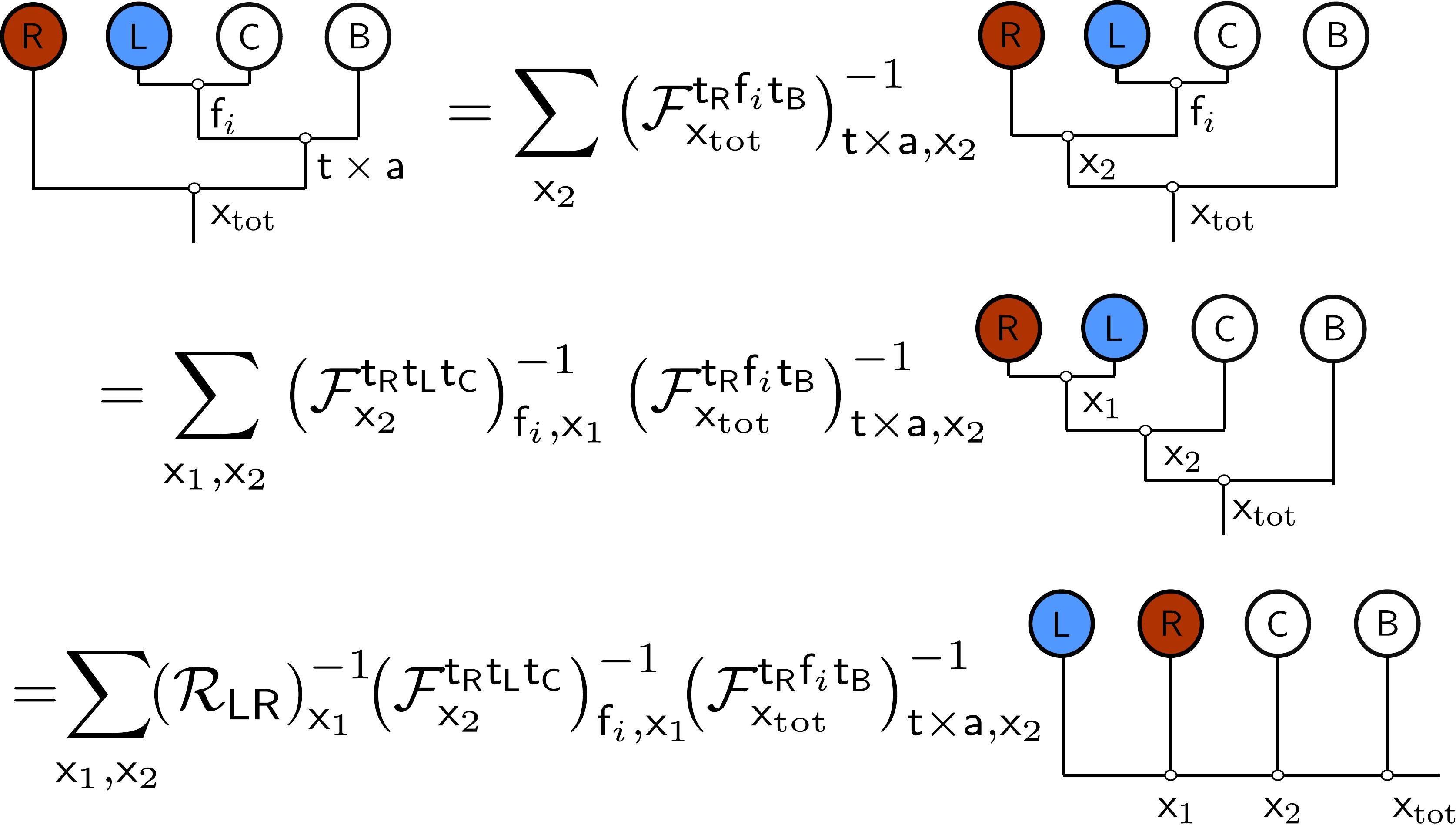}
\caption{The derivation of Eq. \eqref{coefficients}. We transform the ground
  states $\ket{\Psi_{\mathsf{x}_\textrm{tot}}(t)}$ from the basis $\left(
    (\mathsf{t_L} \times \mathsf{t_C} \to \mathsf{f}_i)\times \sf{t_B} \to \sf{t} \times
    \sf{a}\right) $ to the basis \eqref{basis}. The phase factor
  $\left(\mathcal{F}^{\mathsf{t}\mathsf{a}\mathsf{t}}_{\mathsf{x}_\textrm{tot}}
  \right)_{\mathsf{t}\times\mathsf{a},\mathsf{t}\times\mathsf{a}}$ from
  Eq. \eqref{coefficients} is not explicitly
  shown here. \label{fig_basischange}}
\end{figure}

During the first step $0 \leq t \leq T$, the anyon $\mathsf{R}$ is left unpaired from
the other three. The topological charge of the three anyons $\mathsf{L, C, B}$
is then conserved and equal to its initial value
$\mathsf{t_L}\times(\mathsf{t_C}\times\mathsf{t_B})=\mathsf{t}\times\mathsf{a}$. The general
form of a wave function satisfying this constraint is given by:
\begin{equation}\label{Psi_intermediatetimes}
\ket{\Psi_{\mathsf{x}_\textrm{tot}}(t)} = \sum_{\mathsf{x}_1, \mathsf{x}_2,\mathsf{f}_i }  \,U_{\mathsf{x}_\textrm{tot},\mathsf{x}_1, \mathsf{x}_2,\mathsf{f}_i}\, \alpha_{\mathsf{f}_i}(t)\,  \ket{\mathsf{x_1}, {\mathsf{x}_2, \mathsf{x}_\textrm{tot}}},
\end{equation}
where $\alpha_{\mathsf{f}_i}(t)$ can always be chosen to not depend on
$\mathsf{x}_\textrm{tot}$, and the unitary matrix $U$ is the transformation
from the basis $\left( (\mathsf{t_L} \times \mathsf{t_C} \to \mathsf{f}_i)\times \sf{t_B} \to
  \sf{t} \times \sf{a}\right) $, where the anyons \textsf{L}, \textsf{C} and
\textsf{B} fuse directly into $\mathsf{t}\times\mathsf{a}$ before adding the anyon
\textsf{R}, to the basis (\ref{basis}):
\begin{multline}
  U_{\mathsf{x}_\textrm{tot},\mathsf{x}_1, \mathsf{x}_2,\mathsf{f}_i} = \left(\mathcal{F}^{\mathsf{t}\mathsf{a}\mathsf{t}}_{\mathsf{x}_\textrm{tot}} \right)_{\mathsf{t}\times\mathsf{a},\mathsf{t}\times\mathsf{a}} \left( \mathcal{R}_\mathsf{LR}^{-1}\right)_{\mathsf{x}_1}\\
  \times\left( \mathcal{F}^{\sf{t_Rt_Lt_C}}_{\mathsf{x}_{2}}
  \right)_{\mathsf{f}_i,\mathsf{x}_1}^{-1}\left(
    \mathcal{F}^{\sf{t_R}\mathsf{f}_i\sf{t_B}}_{\mathsf{x}_\textrm{tot}}
  \right)_{\sf{t}\times\sf{a},\mathsf{x}_2}^{-1}. \label{coefficients}
\end{multline}
The $\mathcal{F}$- and $\mathcal{R}$-moves required for this transformation are
shown in Fig. \ref{fig_basischange}.

In particular, at $t=0$, only $H_\mathsf{B}\neq 0$ and each
$\ket{\Psi_{\mathsf{x}_\textrm{tot}}(0)}$ is an eigenstate of
$\Pi^\mathsf{B}_\mathsf{a}$ defined in Eq. \eqref{PCB}: 
\begin{multline}\label{Psi0}
  \ket{\Psi_{\mathsf{x_\textrm{tot}}}(0)} =\left(\mathcal{F}^{\mathsf{t}\mathsf{a}\mathsf{t}}_{\mathsf{x}_\textrm{tot}} \right)_{\mathsf{t}\times\mathsf{a},\mathsf{t}\times\mathsf{a}}  \sum_{\mathsf{x}_1,\mathsf{x}_2} \left( \mathcal{R}_\mathsf{LR}^{-1}\right)_{\mathsf{x}_1} \\
  \times\left( \mathcal{F}^{\mathsf{x}_1 \mathsf{t_Ct_B}}_{\mathsf{x}_\textrm{tot}}
  \right)^{-1}_{\mathsf{a},\mathsf{x}_2} \left( \mathcal{F}^{\mathsf{t_R}
      \mathsf{t_L a}}_{\mathsf{x}_\textrm{tot}} \right)^{-1}_{\mathsf{t \times
      a},\mathsf{x}_1} \ket{\mathsf{x}_1,\mathsf{x}_2,\mathsf{x}_\textrm{tot}}\,,
 \end{multline}
These wavefunctions \eqref{Psi0} can be obtained from the Eqs. (\ref{Psi_intermediatetimes}) and (\ref{coefficients}) by substituting $\alpha_{\mathsf{f}_i}(0)=\left(\mathcal{F}^{\sf t_L t_C t_B}_{\sf t \times a} \right)^{-1}_{\sf a, f_i}$ and applying the pentagon equation \cite{bondersonthesis,preskill}. 
The  presence of  the last $\mathcal{F}$  symbol in Eq. \eqref{Psi0} implies  $\mathsf{x}_1  =  \mathsf{x}_\textrm{tot}  \times
 \mathsf{a}$, which simplifies the sum over $\mathsf{x_1}$ due to $\mathsf{a}$ being Abelian.
The phase  factor $(\mathcal{R}_\mathsf{LR}^{-1})_{\mathsf{x}_1}$ is  needed in
 order  to   guarantee  the   independence  of   $\alpha_{\mathsf{f}_i}(t)$  on
 $\mathsf{x}_\textrm{tot}$.  

As $t$ evolves from $0$ to $T$, these states acquire a Berry phase,
\begin{align}\notag
  \theta_{T}&=\int_0^T\bra{\Psi_{\mathsf{x}_\textrm{tot}}(t)}\,\pd_t\ket{\Psi_{\mathsf{x}_\textrm{tot}}(t)}
  \de t = \\
  \label{abelian_phase} &=\int_0^T \sum_{\mathsf{f}_i} \alpha^*_{\mathsf{f}_i} \pd_t
  \alpha_{\mathsf{f}_i} \de t.
\end{align}
The time-independent unitary matrix $U$ naturally drops out of the expression
for the Berry phase. We conclude that the Berry phase acquired in our basis
during the first step is the same for every state, or in other words it is
Abelian.

At $t=T$, only $H_\mathsf{L}\neq 0$, and the ground state wave function must be in
an eigenstate of $\Pi^\mathsf{L}_\mathsf{a}$,
\begin{equation}\label{PsiT}
  \ket{\Psi_{\mathsf{x}_\textrm{tot}}(T)}= \sum_{\mathsf{x}_1}  \left( \mathcal{F}^\mathsf{{t_Rt_Lt_C}}_{\mathsf{x}_{2}} \right)^{-1}_{\mathsf{a},\mathsf{x}_1}  \left( \mathcal{R}_\mathsf{LR}^{-1}\right)_{\mathsf{x}_1} \ket{\mathsf{x}_1,\mathsf{x}_2,\mathsf{x}_\textrm{tot}},
\end{equation}
now with $\mathsf{x}_2=\mathsf{t}\times\mathsf{a}$ since $\mathsf{L}$ and
$\mathsf{C}$ fuse into $\mathsf{a}$, and the phases once again fixed by the
requirement that $\alpha_{\mathsf{f}_i}$ do not depend on
$\mathsf{x}_\textrm{tot}$. Note that the wave functions \eqref{PsiT} are of form
given by Eq. \eqref{Psi_intermediatetimes}. The net result of the evolution
from $t=0$ to $t=T$ is the transfer from $\mathsf{L}$ to $\mathsf{B}$ of an
unpaired topological charge $\mathsf{t}$.

During the second step $T\leq t \leq 2T$ the wave function coefficients can be chosen
to be independent on $\mathsf{x}_\textrm{tot}$ in the basis of
Eq.~\eqref{basis}. The wave function evolves from the eigenstate \eqref{PsiT}
of $\Pi^\mathsf{L}_\mathsf{a}$ into an eigenstate of
$\Pi^\mathsf{R}_\mathsf{a}$. Due to the relation \eqref{symmetry1} and
Eq. \eqref{PsiT} we can write the ground state wave functions at $t=2T$ as
\begin{equation}\label{Psi2T}
  \ket{\Psi_{\mathsf{x}_\textrm{tot}}(2T)}= \sum_{\mathsf{x}_1}  \left( \mathcal{F}^\mathsf{{t_Lt_Rt_C}}_{\mathsf{x}_{2}} \right)^{-1}_{\mathsf{a},\mathsf{x}_1} \ket{\mathsf{x}_1,\mathsf{x}_2,\mathsf{x}_\textrm{tot}}.
\end{equation}
The integral of the Berry connection
$\bra{\Psi_{\mathsf{x}_\textrm{tot}}(t)}\,\pd_t\ket{\Psi_{\mathsf{x}_\textrm{tot}}(t)}$
from $T$ to $2T$ is common to all states and provides an Abelian Berry phase due to the independence of all the coefficients on $\mathsf{x}_\textrm{tot}$.

In the last step, $2T\leq t \leq3T$, we repeat the procedure of the first one. We
write the wave function in a basis $\left( (\mathsf{t_R} \times \mathsf{t_C}
  \to \mathsf{f}_i)\times \sf{t_B} \to \sf{t} \times \sf{a}\right)$, where
$\mathsf{t_R}, \mathsf{t_C}, \mathsf{t_B}$ fuse into
$\mathsf{t}\times\mathsf{a}$ before the anyon $\mathsf{L}$ is added. The
corresponding transformation to the basis \eqref{basis} is given by the
Eq. \eqref{coefficients}, but without the matrix
$\left(\mathcal{R}_{\mathsf{LR}}\right)^{-1}$. This ensures that the wave
function $\ket{\Psi_{\mathsf{x}_\textrm{tot}}(t)}$ stays continuous at $t=2T$. In
this last step, the wave function acquires another Abelian Berry phase and ends
up again in an eigenstate of $\Pi^\mathsf{B}_\mathsf{a}$. We end up with:
\begin{multline}\label{Psi3T}
  \ket{\Psi_{\mathsf{x_\textrm{tot}}}(3T)}=\left(\mathcal{F}^{\mathsf{t}\mathsf{a}\mathsf{t}}_{\mathsf{x}_\textrm{tot}}
  \right)_{\mathsf{t}\times\mathsf{a},\mathsf{t}\times\mathsf{a}}
  \sum_{\mathsf{x}_1,\mathsf{x}_2} \left( \mathcal{F}^{\mathsf{x}_1
      \mathsf{t_Ct_B}}_{\mathsf{x}_\textrm{tot}}
  \right)^{-1}_{\mathsf{a},\mathsf{x}_2} \\\times\left( \mathcal{F}^{\mathsf{t_R}
      \mathsf{t_L a}}_{\mathsf{x}_\textrm{tot}} \right)^{-1}_{\mathsf{t \times
      a},\mathsf{x}_1} \ket{\mathsf{x}_1,\mathsf{x}_2,\mathsf{x}_\textrm{tot}}.
\end{multline}

\begin{figure*}[t!]
 \centering
 \includegraphics[width=\textwidth]{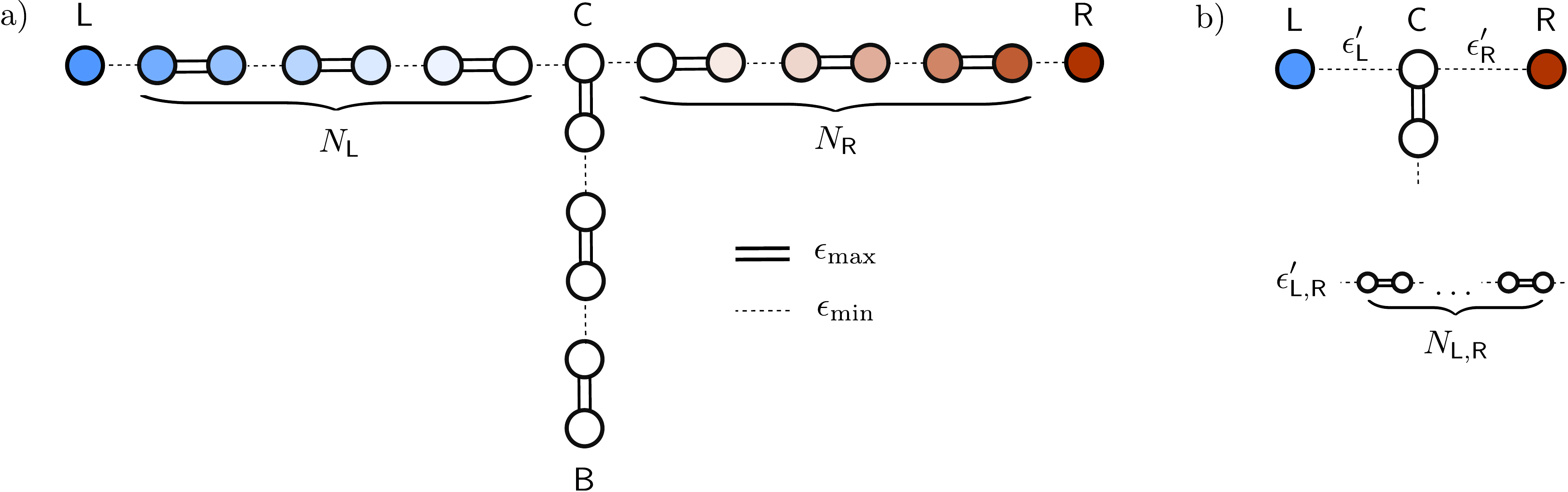}
 \caption{Panel (a): three staggered anyon chains forming a T-junction. Weak
   ($\epsilon_\textrm{min}$, dashed lines) and strong ($\epsilon_\textrm{max}$, double solid
   lines) couplings alternate. The bottom arm of the T-junction, connecting the
   original anyons $\mathsf{C}$ and $\mathsf{B}$, is in a dimerized phase with
   no unpaired anyons and approximately contains no net topological charge. On
   the other hand, in the right and left arm the dimerization leaves two almost
   unpaired anyons $\mathsf{L}$ and $\mathsf{R}$ at the end (blue and red
   dot). Due to the residual coupling, the topological charge of $\mathsf{L}$
   and $\mathsf{R}$ is spread over the neighboring anyon pairs, as represented
   by the color gradings. The left and right arm are in therefore in the
   non-trivial phase. The two arms interact weakly via the centre of the
   T-junction, leading to renormalized couplings $\epsilon'_\mathsf{L}$ and
   $\epsilon'_\mathsf{R}$ between $\mathsf{L}$, $\mathsf{R}$ and $\mathsf{C}$, as in
   panel (b). The residual interaction splits the ground state degeneracy of an
   energy exponentialy small in the length of the chains.}
\label{chains}
\end{figure*}

Having performed an adiabatic evolution over a closed path, the final wave
function must be connected to the initial one via a unitary matrix
$\mathcal{U}$, $\ket{\Psi(3T)}=\mathcal{U}\ket{\Psi(0)}$. Using Eq. \eqref{Psi0} and
\eqref{Psi3T} we find
\begin{equation}
  \bracket{\Psi_{\mathsf{x}_\textrm{tot}}(0)}{\Psi_{\mathsf{x}_\textrm{tot}}(3T)}=(\mathcal{R}_{LR})_{\mathsf{x}_1}
\end{equation}
where we recall that $\mathsf{x}_\textrm{tot}=\mathsf{x}_1\times \mathsf{a}$. For the
whole wave function we can write
\begin{equation}\label{result}
\ket{\Psi(3T)}=\mathcal{R}_\mathsf{LR}\,\ket{\Psi(0)}
\end{equation}
up to an Abelian Berry phase. This means that the braiding of anyons
$\mathsf{L}$ and $\mathsf{R}$ was performed in the adiabatic cycle. By performing the whole protocol in reverse, we obtain instead the inverse braiding.

\section{Discussion and conclusions}\label{Sec_conclusions}

\subsection{Restoring scalability and topological protection}\label{Sec_chains}

The braiding procedure of Sec.~\ref{Sec_cycle} relies on the ability to turn
off the pairwise interactions $H_\textsf{K}$ completely. This is only possible
if the separation between the anyons becomes infinite, and hence one may argue
that this procedure is only approximating topological quantum computation. In a
finite system the non-Abelian Berry phase will in general have a correction,
and additionally non-adiabatic errors will appear due to the presence of finite
ground state splitting \cite{dassarma11}.

This imperfection can be removed and the topological nature of the braiding can
be restored by bringing the anyons $\mathsf{L}, \mathsf{R}, \mathsf{B}$ further
away from the central one $\mathsf{C}$. If anyonic chains with controllable
couplings are then introduced along the three arms of the T-junction (see
Fig. \ref{chains}), this still allows to perform the braiding in a similar
fashion, but with a higher fidelity. Since we are interested in the low energy
spectrum of the Hamiltonian we approximate the interactions between
nearest-neighbor anyons $\mathsf{K},\mathsf{K'}$ with the projector
$\Pi_\mathsf{a}^{\mathsf{K,K'}}$ over their lowest energy topological charge and
we consider all the other fusion channels to have the same energy, so that the
Hamiltonian of each junction becomes:
\begin{equation}
H_\mathsf{K,K'}= -\epsilon \Pi_\mathsf{a}^\mathsf{K,K'},
\end{equation}
where $\mathsf{a}$ should again be Abelian. We require that $\epsilon$ can be varied
in a range $(\epsilon_\textrm{min}, \epsilon_\textrm{max})$, so that the chains can be driven
into a staggered phase with alternating weak and strong couplings, as in the
Kitaev Majorana chain \cite{Kit01} and its parafermionic generalization
\cite{fendley12}.

The termination of the chain ending with a weak link differs from the
termination by a strong link by the presence of an extra $\mathsf{t}$ anyon,
and the chain ending with a strong link can be continuously connected to a
chain of fully fused $\mathsf{a}$-type anyons. This means that if the chain is
gapped, whenever it ends in a weak link, its end has a topological charge of
$\mathsf{t}$, spread over several anyons, as shown in Fig.~\ref{chains}. While
we are not aware of a proof that a general anyonic chain with staggered
antiferromagnetic couplings is gapped, it is true for many relevant cases
\cite{gils09,fid08,lau12}. When $\epsilon_\textrm{min} \ll \epsilon_\textrm{max}$, the
effective minimal coupling between an unpaired anyon at the edge of the
T-junction and the central anyon $\mathsf{C}$ can be calculated perturbatively,
and it is equal to $\epsilon'\simeq \epsilon_\textrm{min}
(\kappa\epsilon_\textrm{min}/\epsilon_\textrm{max})^N$, with $N$ the number of anyon pairs in the
chain, and $\kappa$ a geometric factor which depends on the specific anyon
model. For Ising anyons $\kappa = 1$, and for Fibonacci anyons $\kappa=2/\phi^2$, with
$\phi=\left( 1+\sqrt{5}\right)/2$ the golden ratio \cite{fid08,lau12}. The maximal
coupling is achieved in the staggered configuration which ends with a strong
bond, and the maximal coupling $\epsilon_\textrm{max}$ is only weakly modified.

\begin{figure}[tb]
\includegraphics[width=\linewidth]{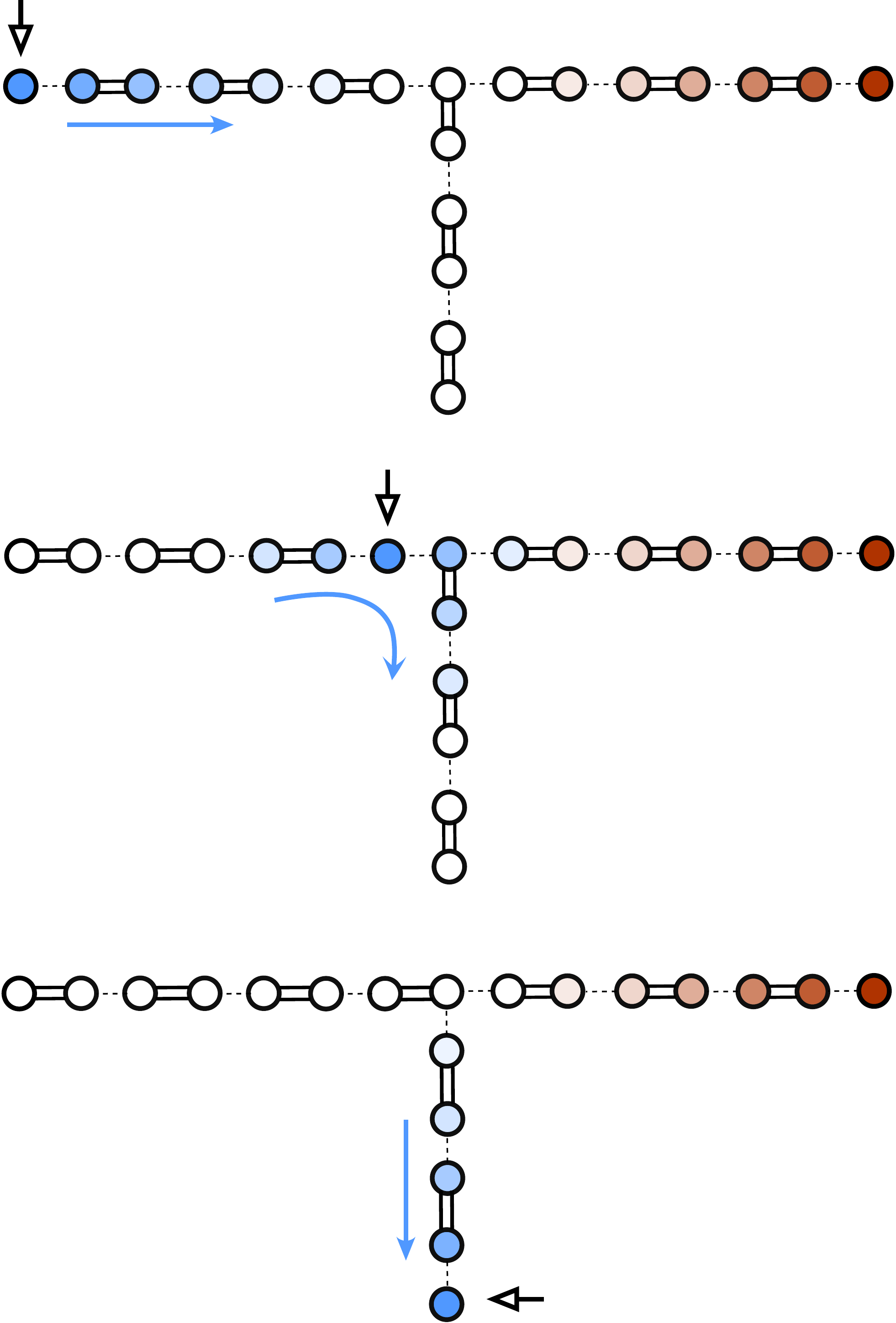}
\caption{The first step of the adiabatic braiding sequence realized in a system
  of staggered anyonic chains. The topological charge $\mathsf{t}$ is moved
  from the left arm of the junction to the bottom arm. As in Fig. \ref{chains},
  blue and red colors represent a topological charge $\mathsf{t}$ spread over
  several anyons. The charges are localized at domain walls between the two
  possible phases of the staggered chain. Domain walls can be moved: each
  movement involves three different anyons of the chain. The domain wall that
  is moved is marked by a black arrow.\label{fig_step}}
\end{figure}

To implement the braiding, each part of the adiabatic evolution can be
decomposed into steps which require to change the pairwise couplings of three
anyons, just as it happens for the steps illustrated in
Fig. \ref{fig_cycle}. In this way, during the adiabatic cycle, we create and
move domain walls which drive the transition between the two different
staggered configurations of the chains (see Fig. \ref{fig_step}). The two
unpaired topological charges encoding the computational degree of freedom are
localized in these domain walls which are moved along the three arms. Since the
distance between the unpaired charges is always larger than the length $N$ of a
single arm of the T-junction, their residual interaction is exponentially
suppressed, allowing to likewise exponentially suppress the error in the final
result.

\subsection{Summary}

In summary, we have investigated an approach to topological quantum
computation. In order to implement the necessary braiding operations of
non-Abelian anyons, we couple the anyons instead of moving them or measuring
their state. We have considered a simple system composed of four interacting
non-Abelian anyons in a T-junction geometry and we have shown how adiabatic
control over the interactions results in the Berry matrix expected when two
anyons are moved around each other. If the coupling between the anyons cannot
be completely turned off, errors are introduced in the braiding operations due
to the residual splitting of the ground state degeneracy. We have discussed how
these errors can be limited by means of enlarging the number of anyons involved
in the adiabatic evolution. The protection is exponential in the number of
anyons which are added to the system, so the whole procedure is similar to
increasing the separation between anyons in the original approach.

Our approach, inspired by recent theoretical proposals for the braiding of
Majorana fermions in superconductors, is applicable to most anyon models. These
include all the SU$(2)_{k}$ models (such as the Ising and Fibonacci anyons
expected to appear in fractional quantum Hall systems), as well as the
fractionalized Majorana fermions very recently proposed in
Refs. \cite{cla12,lin12,cheng12,vaezi12,wen12,fendley12}.

A possible implementation of our scheme in the fractional quantum Hall systems,
would require to engineer systems of dots hosting single anyonic quasiparticles
and to tune their interactions via the voltages induced by gates or scanning
tips, in a similar spirit to the blockade measurement of topological charge
\cite{vanheck12}.

Alternative, but even more exotic, implementations of this scheme include for
example the braiding procedure presented in Refs. \cite{lin12,barkeshli12} for
fractional Majorana fermions in superconductor/quantum-Hall
heterostructures. Additionally, the recent progress in the design of several
systems thought to host non-Abelian excitations, ranging from physical
realizations of the Kitaev honeycomb lattice model \cite{kitaev06} (see, for
example \cite{duan03,zoller06,jackeli09}) to ultracold atomic gases subjected
to artificial gauge potentials \cite{cooper08,lewensteinbook}, could also fall
into the category of systems where interactions between anyons are easier to
control than their positions.

\section*{Acknowledgements.}

We thank Parsa Bonderson for pointing out a mistake in Eq. \eqref{coefficients}
appearing in a previous version of this manuscript, and for helpful
comments. This project was supported by the Dutch Science Foundation NWO/FOM
and a ERC Advanced Investigator Grant. AA was partially supported by a Lawrence
Golub Fellowship.


\begin{thebibliography}{25}
\bibitem{isingnote}
In the general case, the interactions between $\mathsf{C}$ and the other anyons may favor different Abelian fusion channels $\mathsf{a_K}$ if all the fusions $\mathsf{t} \times \mathsf{a_K}$ assume the same topological charge.
The main example is the case of Ising anyons $\sigma$, which obey the fusion rules $\sigma \times \sigma=\mathbb{I} + \Psi$, $\sigma \times \mathbb{I} = \sigma \times \Psi = \sigma$, with both $\mathbb{I}$ and $\Psi$ Abelian; in this case changing the favored fusion channel of the pairwise interaction between $\mathsf{L}$ (or $\mathsf{R}$) and $\mathsf{C}$ determines a change in the chirality of the braiding \cite{Sau11, hal12}. This is due to an additional symmetry of the Ising anyon model:
\begin{equation}
 \mathcal{R}^{-1}_\mathsf{LR} \Pi^\mathsf{L}_\Psi \mathcal{R}_\mathsf{LR} = \mathcal{R}_\mathsf{LR} \Pi^\mathsf{L}_\mathbb{I} \mathcal{R}^{-1}_\mathsf{LR}
\end{equation}
where $\Pi_\Psi=1-\Pi_\mathbb{I}$ is the projector over the fermionic fusion channel $\Psi$.  This relation implies that changing the favored fusion channel for one of the two anyons $\mathsf{R}$ or $\mathsf{L}$, the role of $\mathcal{R}_\mathsf{LR}$ and $\mathcal{R}^{-1}_\mathsf{LR}$ are exchanged throughout the adiabatic cycle and effectively reverses the braiding direction.
\end{thebibliography}
\end{document}